\documentclass[final,3p,times,twocolumn]{elsarticle}
\usepackage{amssymb}
\usepackage[intlimits]{amsmath}
\usepackage{amsfonts}
\usepackage{subfigure}
\usepackage[colorlinks]{hyperref}
\usepackage{graphicx}
\usepackage[usenames]{color}

\newcommand{\sh}[1]{#1\hskip-7pt \diagup}

\newcommand{\tr}{\textrm{tr}}

\journal{Physics Letters B}

\begin{document}

\begin{frontmatter}

\title{Leading-order calculation of hadronic contributions to the muon $g-2$ using the Dyson-Schwinger approach}

\author[Giessen]{Tobias Goecke}
\author[Giessen,GSI]{Christian S. Fischer}
\author[Madrid]{Richard Williams}
\address[Giessen]{Institut f\"ur Theoretische Physik, 
 Universit\"at Giessen, 35392 Giessen, Germany}
\address[GSI]{Gesellschaft f\"ur Schwerionenforschung mbH, 
  Planckstr. 1  D-64291 Darmstadt, Germany}
\address[Madrid]{Dept. F\'isica Teor\'ica I, Universidad Complutense, 28040 Madrid, Spain}  
  
\begin{abstract}
We present a calculation of the hadronic vacuum polarization (HVP)
tensor within the framework of Dyson--Schwinger equations. To this end
we use a well-established phenomenological model for the quark-gluon
interaction with parameters fixed to reproduce hadronic observables. 
From the HVP tensor we compute both the Adler function and the HVP 
contribution to the anomalous magnetic moment of the muon, $a_\mu$. 
We find $a_\mu^{HVP}= 6760\times 10^{-11}$  which deviates about two 
percent from the value extracted from experiment. Additionally, we make
comparison with a recent lattice determination of $a_\mu^{HVP}$ and find
good agreement within our approach.  We also discuss
the implications of our result for a corresponding calculation of the 
hadronic light-by-light scattering contribution to $a_\mu$. 
\end{abstract}

\begin{keyword}


\end{keyword}

\end{frontmatter}

%
%
%
%

\section{Introduction\label{intro}}
One of the most interesting places to search for new physics beyond the
Standard Model (SM) is the anomalous magnetic moment of the muon, $a_\mu$. 
It is dominated by QED effects, however due to the heavy mass of the muon it 
is also sensitive to other corrections. Aside from weak interaction
contributions which can be evaluated in perturbation theory, one also 
has to include effects from QCD. Since the latter are intrinsically
non-perturbative at the scales relevant to the calculation, they are much 
harder to include systematically.

\begin{figure}[t!]
		\centering	\subfigure[][]{\label{fig:hadroniclo}\includegraphics[width=0.30\columnwidth]{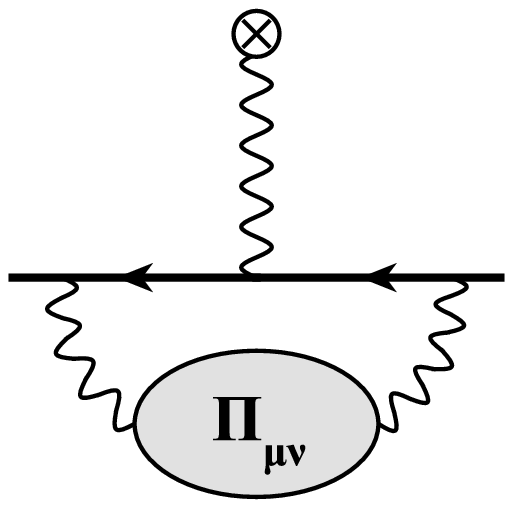}}
	\hspace{0.08\columnwidth}
	\subfigure[][]{\label{fig:hadroniclbl}\includegraphics[width=0.30\columnwidth]{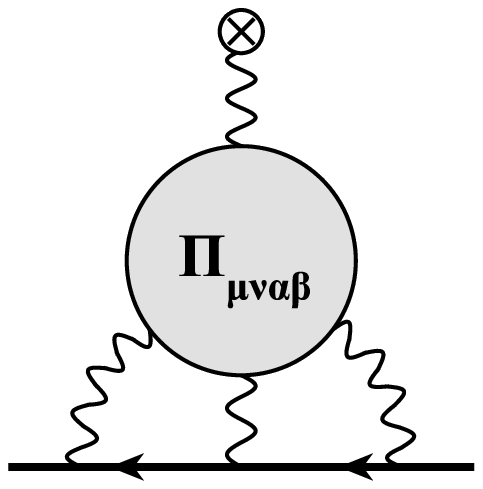}}
      \caption{The two classifications of corrections to the photon-muon
	vertex function: (a) hadronic vacuum polarization contribution to $a_\mu$. The vertex is 
               dressed by the vacuum polarization tensor $\Pi_{\mu\nu}$;
		   (b) the hadronic light-by-light scattering contribution to
		   $a_\mu$.}
\end{figure}

Experimental efforts at Brookhaven National Lab and elaborated theoretical 
efforts of the past ten years have pinned down $a_\mu$ to the $10^{-11}$ level, 
leading to significant deviations between theory \cite{Jegerlehner:2009ry} 
and experiment \cite{Bennett:2006fi,Roberts:2010cj}:
      \begin{align}
		\label{eqn:amuexperiment}
            \mbox{Experiment:} \,\,\,\,
			&116\,592\,089.0(63.0)\times 10^{-11} \;\; , 
		\\
		\label{eqn:amutheoretical}
            \mbox{\phantom{wwu}} \mbox{Theory:} \,\,\,\,
			&116\,591\,790.0(64.6)\times 10^{-11} \;\; .
      \end{align}
This $3.3\,\sigma$ deviation might be seen as a sign for new physics,
however confirmation requires that the uncertainties of both theory and 
experiment must be reduced yet further. The error on the theoretical side 
is dominated by hadronic contributions involving non-perturbative QCD
dynamics. The leading order hadronic contribution is the hadronic vacuum 
polarization insertion (HVP) shown in Fig.~\ref{fig:hadroniclo}.
At present, this diagram also dominates the theoretical error of $a_\mu$
from the SM. One obtains~\cite{Jegerlehner:2008zza}
      \begin{align}\label{eqn:hadroniclo}
		a_\mu^{(\mathrm{HVP})} =    [6\,903.0(52.6) - 100.3(1.1)]\times 10^{-11} \;\; ,
      \end{align}
for the leading and subleading contributions, see also 
Refs.~\cite{Davier:2010nc,Hagiwara:2011af} for recent updates.
The HVP-tensor ($\Pi_{\mu\nu}$) involved in the calculation of the
leading order result can be obtained from experimental input by recourse
to the optical theorem; such results can then be regarded as being
model independent. However, note that models may be involved
in the analysis or extraction of this experimental data, especially in the
(dominant) low $Q^2$ region~\cite{Benayoun:2011mm}.

The diagram that, in the literature, yields the second largest theoretical error is that of 
hadronic light-by-light scattering (LBL),~Fig. \ref{fig:hadroniclbl}.
It is extremely difficult to measure and therefore needs to be determined
from theory alone. There is a long history
of different approaches to this problem, see Ref.~\cite{Jegerlehner:2009ry}
for an overview. Recently, we provided a re-evaluation of $a_\mu^{LBL}$
in the framework of Dyson-Schwinger and Bethe-Salpeter equations
of QCD \cite{Fischer:2010iz,Goecke:2010if}. Starting with a phenomenologically 
successful model for the quark-gluon interaction, we determined dynamically 
the momentum dependent quark propagator, the corresponding meson Bethe-Salpeter
amplitudes and the quark-photon vertex and used these as building blocks
for our calculation of $a_\mu^{LBL}$. In contrast to previous approaches, we 
automatically included effects in the quark-photon interaction that are 
induced by gauge invariance. This can be seen as one of the improvements
that DSEs have over typical effective approaches to QCD. Our results indicate that 
the theoretical value of Eq.~(\ref{eqn:amutheoretical}) may indeed receive 
additional positive contributions that reduce the discrepancy with experiment. 
The precise size of these contributions, however, will only become clear 
once we reduce the approximations made in~\cite{Fischer:2010iz,Goecke:2010if}. 

While work in this direction is in progress, we find it prudent to 
elucidate upon and justify our approach via a calculation of
the hadronic vacuum polarization $\Pi_{\mu\nu}$. Although this quantity 
in principle need not be determined from theory for the purposes of $a_\mu$, 
it serves as an important testing ground for any approach used for 
calculating hadronic contributions to $a_\mu$
\cite{deRafael:1993za,Pallante:1994ee,Bell:1996md,Dorokhov:2004ze,DellaMorte:2010sw}. 
In this letter we provide results for the HVP
contribution to the muon anomaly together with the Adler function. We
employ the same model and philosophy as used recently in our calculation
of hadronic light-by-light scattering~\cite{Fischer:2010iz,Goecke:2010if}.
By comparing to the results extracted from experiment and to recent
lattice calculations \cite{Feng:2011zk} we will demonstrate that our 
approach provides meaningful and quantitatively reliable results. We also 
believe that our results serve to address and invalidate an argument made 
by the authors of Ref.~\cite{Boughezal:2011vw}. There, one-loop radiative 
corrections to $a_\mu^{HVP}$ and $a_\mu^{LBL}$ in a constituent quark
model have been invoked to argue against large effects from vertex 
corrections. 
While their calculation is no doubt correct -- within
the limitations of using perturbation theory at strong coupling scales -- 
the relevance of their results to the case of g-2 seems rather limited.
This will be discussed in more detail below.

The outline of the letter is as follows. In section~\ref{sec:HVPTensor} 
we will introduce the hadronic vacuum polarization, starting with its
basic definition and its calculation within the functional approach. 
In section \ref{sec:Framework} we present the framework that we employ 
in this paper, the Dyson--Schwinger (DSE) and Bethe-Salpeter (BSE) 
equations. This is followed by our results and a discussion pertaining 
$a_\mu^{HVP}$ and the Adler function in section~\ref{sec:Results}. Finally we 
summarize and discuss the relevance of our results for $a_\mu^{LBL}$ in
the concluding sections.

%
%
%
%

\section{The Hadronic Vacuum Polarization Contribution\label{sec:HVPTensor}}
In the following we give the basic definitions
concerning the HVP tensor, the muon anomaly and the Adler function.
Throughout this work we will employ Euclidean space conventions.

\subsection{Basic Definitions\label{sec:BasicDefs}}
The hadronic vacuum polarisation tensor $\Pi_{\mu\nu}$ is defined 
as that part of the one particle irreducible (1PI) photon self
energy that is generated by QCD dynamics. It can be obtained from the 
photon Dyson-Schwinger equation
\begin{align}
  D^{-1}_{\mu\nu}(q) = Z_3\, (D^{(0)}_{\mu\nu}(q))^{-1}  - e^2\, \Pi_{\mu\nu}(q) \;\; ,
  \label{eqn:photonDSE}
\end{align}
where $D_{\mu\nu}$ is the full photon propagator,
$D^{(0)}_{\mu\nu}$ the bare propagator and $Z_3$
is the photon renormalisation constant. The hadronic tensor $\Pi_{\mu\nu}$, 
specified explicitly below, can also be seen as the 1PI-part of the current correlator
\begin{align}
  \Pi_{\mu\nu}(q) = \int_x\,\,e^{i\,q\cdot x} \langle j_\mu(x)j_\nu(0) \rangle_{\textrm{1PI,hadr.}} \;\; ,
  \label{eqn:HVPTensorDef}
\end{align}
with $\int_x=\int d^4x$ and the electromagnetic quark current $j_\mu$ given by
      \begin{align}  \label{eqn:quarkcurrent}
            j_\mu &= \frac{2}{3}\bar{u}\gamma_\mu u 
                   - \frac{1}{3}\bar{d}\gamma_\mu d 
                   - \frac{1}{3}\bar{s}\gamma_\mu s
                   + \frac{2}{3}\bar{c}\gamma_\mu c
                   - \frac{1}{3}\bar{b}\gamma_\mu b\;\;.
      \end{align}
Here $u$, $d$, $s$, $c$ and $b$ are the respective quark spinors.
It follows from the Ward Takahashi identity (WTI) $q_\mu\Pi_{\mu\nu}=0$
that the HVP tensor is transverse:
\begin{align}
  \Pi_{\mu\nu}(q) = \left(\delta_{\mu\nu}-\frac{q_\mu q_\nu}{q^2}\right)
  \,q^2\, \Pi(q^2) \;\; ,
  \label{eqn:HVPScalarDef}
\end{align}
which serves as a definition of the scalar vacuum polarization $\Pi(q^2)$.
The quantity $\Pi(q^2)$ is logarithmically divergent
and has to be renormalized. 
We choose the condition $\Pi(0)=0$ which
leaves the definition of the electric charge intact. More details
concerning our renormalization prescription can be found below. 

Once we have obtained the renormalized HVP scalar,
$\Pi_\mathrm{R}(q^2)$, the leading hadronic contribution to $a_\mu^{HVP}$
can be calculated~\cite{deRafael:1993za}
\begin{align}
  a_\mu^{\mathrm{HVP}}= \frac{\alpha}{\pi}\int_0^1\!\! dx\,\,(1-x) \left[-e^2\Pi_\mathrm{R}\left(\frac{x^2}{1-x}m_\mu^2  \right)  \right] \;\; ,
  \label{eqn:anomalyIntegral}
\end{align}
where $m_\mu$ is the muon mass and $\alpha=e^2/4\pi$ is the fine
structure constant.

The Adler function $D(q)$ is defined as the logarithmic derivative 
of the polarization scalar
\begin{align}
  D(q) = - q^2 \frac{d\,\Pi(q^2)}{d\,q^2} \;\; .
  \label{eqn:AdlerFunctionDef}
\end{align}
The HVP tensor and the Adler function can be obtained independently 
of the model from experiment, exploiting dispersion relations
(see e.g. \cite{Jegerlehner:2009ry,Jegerlehner:2008zza} for details).

\subsection{Expansion in a functional approach}
\label{sec:FunctionalExpansion}
\begin{figure}[t]
\begin{align}
  %
  \parbox{2.3cm}{\includegraphics[height=0.03\textheight]{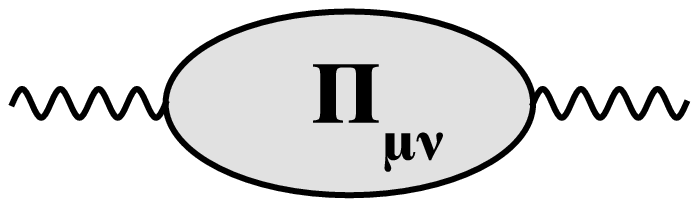}} &=
  \parbox{2.4cm}{\includegraphics[height=0.05\textheight]{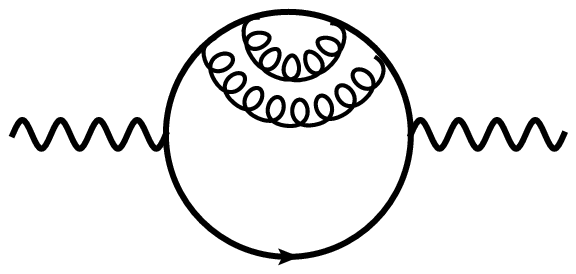}}+
  \parbox{2.4cm}{\includegraphics[height=0.05\textheight]{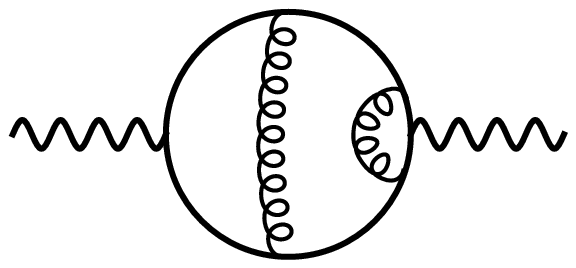}}\notag\\
  &+\parbox{2.4cm}{\includegraphics[height=0.05\textheight]{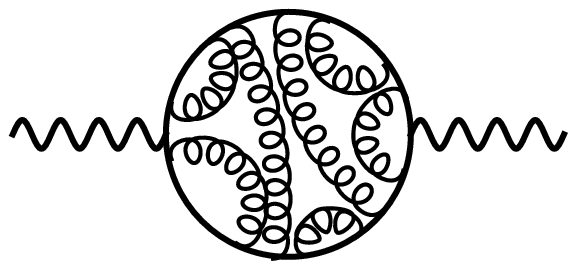}}+\cdots
  \notag \\
  &=
  \parbox{2.4cm}{\includegraphics[height=0.06\textheight]{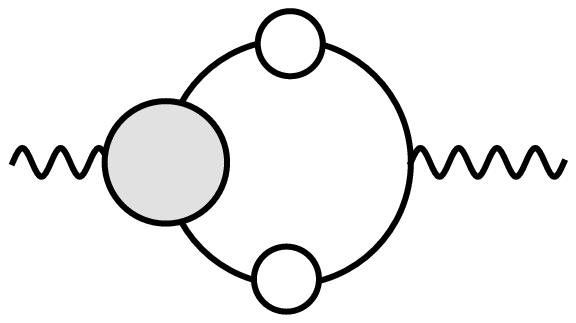}}\,. \notag
\end{align}
\caption{The photon vacuum polarization and its expansion in planar diagrams.}
\label{eqn:expandTwopointFkt}
\end{figure}

In a functional approach the vacuum polarisation tensor is essentially
the photon self-energy. For hadronic contributions these photons couple
to quarks, which in turn couple to gluons. It thus contains a
resummation of an infinity of diagrams. In the spirit of the $1/N_c$
counting we consider only those diagrams which are planar. This infinite subset of 
diagrams is the same as those considered in~\cite{Fischer:2010iz,Goecke:2010if}.
The resulting expansion is depicted graphically in Fig.~\ref{eqn:expandTwopointFkt}.
The first diagram on the right hand side shows gluonic
corrections that non-perturbatively dress the current quark. The second
diagram shows gluonic corrections to the quark-photon vertex. Both
classes of diagrams are indicated in the third diagram, showing the
complexity of the resummation. These are finally written in terms of
fully-dressed one-particle irreducible Green's functions (propagators and
vertices marked by circles) in the second line of the equation. 
These are calculated self-consistently within a rainbow-ladder 
approximation to their DSEs, detailed in the next section.  Note
that the diagram in the last line of Fig.~\ref{eqn:expandTwopointFkt} 
is an exact representation of the hadronic tensor. The truncation 
takes place on the level of the propagator and the vertex.

\section{Framework}
\label{sec:Framework}

In the following we summarize the calculation scheme employed in this
paper; more explicit details can be found in Ref.~\cite{Goecke:2010if}.
The Dyson--Schwinger equations (DSEs) are exact relations
amongst the Green's function of a given theory. Since
they constitute an infinite tower of coupled integral equations
a truncation has to be employed to provide tractability. For
the calculation of the Adler function and the muon anomaly, we need the 
quark propagator, quark-photon vertex and hadronic vacuum polarisation
tensor. These are obtained from their respective DSEs, which we detail below.

\subsection{The Quark DSE}
\begin{figure}[t]
      \includegraphics[width=0.97\columnwidth]{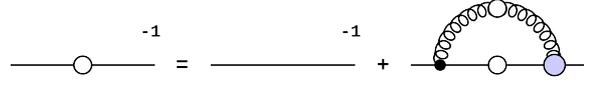}
      \caption{Dyson--Schwinger equation for the quark propagator. Specification
               of the fully-dressed gluon propagator (wiggley line) and 
	       quark-gluon vertex (grey blob) defines the truncation 
               scheme.}\label{fig:quarkdse}
\end{figure}

We begin with the dressed quark propagator $S(p)$, 
	\begin{align}
		S(p) =  Z_f(p^2) \left(i \sh{p} + M(p^2)\right)^{-1} \;\;, \label{eqn:quark}
	\end{align}
which is characterized by the momentum dependent quark mass function $M(p^2)$ 
and the wave function 
$Z_f(p^2)$. These are obtained as a solution of the quark DSE given 
diagrammatically in Fig.~\ref{fig:quarkdse}. On the right hand side the
inverse bare quark propagator is given by 
$S^{-1}(p) = Z_2 \left(i \sh{p} + m\right)$ with quark
renormalization factor $Z_2$ and the bare mass $m$.
The quark self-energy contains the gluon propagator, given in Landau
gauge as
	\begin{align}
		D_{\mu\nu}(k) =  \left( \delta_{\mu\nu}-\frac{k_\mu k_\nu}{k^2}  
		\right)\frac{Z(k^2)}{k^2}\;\;, \label{eqn:gluon}
	\end{align}
with dressing function $Z(k^2)$. In addition the dressed quark-gluon vertex 
$\Gamma_\mu(p,q)$ is required. A simple, yet phenomenologically successful
approximation of the quark-gluon interaction has been suggested by Maris and 
Tandy~\cite{Maris:1999nt}. Here only the leading Dirac structure of the vertex 
is retained $\Gamma_\mu(k^2)=\gamma_\mu \Gamma^{\mathrm{YM}}(k^2)$ 
and the dressing of the Yang-Mills (YM) part of the vertex is chosen to depend on the gluon momentum $k$ only. The 
combination of the gluon- and vertex-dressing functions is then modeled as
	\begin{eqnarray}
\hspace*{-1.5mm}		Z(k^2) \Gamma^{\mathrm{YM}}(k^2) &=&  \frac{4\pi}{g^2} 
      	 \bigg( \frac{\pi}{\omega^6}D k^4 \exp(-k^2/\omega^2)\label{eqn:maristandy}\\
	 &&\hspace*{-20mm}+\frac{2\pi \gamma_m}{\log\left( \tau+\left(1+k^2/\Lambda_\mathrm{QCD} \right)^2\right)}
		 \left[ 1-e^{-k^2/\left( 4m_t^2 \right)} \right]\bigg)\;\;,\nonumber 
	\end{eqnarray}
with $m_t= 0.5\,{\rm GeV}$, $\tau\;=\;\mathrm{e}^2-1$, $\gamma_m=12/(33-2N_f)$, 
$\Lambda_\mathrm{QCD} \;=\;0.234\,{\rm GeV}$, $\omega=0.4\,{\rm GeV}$ and $D=0.93\,{\rm GeV}^2$. 
This model interaction assumes the form of the one loop running coupling 
of QCD at momenta $k^2 >> \Lambda^2_\mathrm{QCD}$ and provides enough 
interaction strength in the infrared for dynamical chiral symmetry breaking 
to occur. 

Combining the DSE with the corresponding Bethe-Salpeter equation (BSE) one can 
determine mesonic bound state masses and their decay constants. The model parameters 
$\omega$ and $D$ are then chosen such that the physical value of the pion decay 
constant is reproduced. The quark masses have then been fixed by comparison with 
experimental meson masses in the pseudoscalar meson sector, cf. the first set in
Table~\ref{tab:param}. Together with the self-consistently 
calculated quark-photon vertex (see below) electromagnetic properties 
such as electromagnetic form factors and charge radii can be obtained 
\cite{Maris:1999bh,Bhagwat:2006pu} that are in good agreement with 
experiment. This is also true for heavy flavors as discussed in 
Ref.~\cite{Maris:2006ea}. Especially important for the calculation of
the HVP tensor is, however, the behavior of the model in the vector
meson channel. Here, the deviation to experiment is on the five percent 
level, as can be seen from the first line of Table~\ref{tab:param}. 
It is therefore not unreasonable to expect that the model provides a 
good description of hadronic contributions to $\Pi_{\mu\nu}$ up to 
potential deviations of the order of five to ten percent to the experimental 
value. One possibility to investigate the systematic error of the
model further, is to fix the bare quark masses not with pseudoscalar meson
masses, but with the vector meson sector. The corresponding values are
given in the second line of Table~\ref{tab:param}. Naturally, this is 
at the expense of the pseudoscalar sector, which reacts quadratically
to a change in the mass parameters, as opposed to the linear change of 
the vector meson sector. Below, we will employ both mass parameter sets in our 
calculation of $a_\mu^{\textrm{HVP}}$ and estimate the model inherent 
systematic error by a comparison of the results.

\begin{table}[t]
  \centering
  \begin{tabular}{c|c|c|c|c|c|c}
    [MeV]	&	$m_{u,d}$	&	$m_{s}$ 	& 	$m_{\pi}$ & $m_{K}$ & $m_{\rho}$ & $m_{\phi}$	\\\hline\hline
	set I	&    $3.7$  	&	$85$ 		&	$138$	  &	$495$	&	$740$		 &	$1080$	\\\hline
	set II	&    $11$  		&	$72$ 		&	$240$	  &	$477$   &	$770$		 &	$1020$	\\\hline
  \end{tabular}
  \caption{Two choices for the light bare quark masses at $\mu^2=(19\,\mbox{GeV})^2$ and the resulting meson 
  masses (in MeV) in the pseudoscalar and vector meson sector. For the heavy quarks we always take $m_c = 827$ MeV and 
  $m_b = 3680$ MeV which lead to good results for charmonia and bottomonia in the pseudoscalar and vector channel.}
  \label{tab:param}
\end{table}

\subsection{The quark-photon vertex}
The second ingredient necessary for the determination
of the hadronic tensor through Eq.~(\ref{eqn:expandTwopointFkt})
is the fully dressed quark-photon vertex.
This quantity is obtained self-consistently from its inhomogeneous Bethe-Salpeter equation
\begin{align}
  \Gamma_\mu(P,k)& = Z_2\,\gamma_\mu  +   \frac{4}{3}\,g^2\, Z_2^2\,  \notag\\
  \times \int_q &\left[\gamma_\alpha S(q_-) \Gamma_\mu(P,q)S(q_+)\gamma_\beta\right]
  D_{\alpha\beta}( r^2)   \Gamma^{\mathrm{YM}}( r^2) \;\;,
  \label{eqn:quarkphotondse}
\end{align}
where $r=q-k$ and $\int_q=\int\frac{d^4q}{(2 \pi)^4}$.%
\begin{figure}[b!]
      \includegraphics[width=0.95\columnwidth]{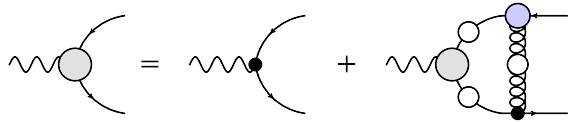}
      \caption{Inhomogeneous BSE for the quark-photon vertex.}\label{fig:quarkphotondse}
\end{figure}
We show the BSE pictorially in Fig. \ref{fig:quarkphotondse}. Again we use the 
ladder truncation ensuring that both the axial-vector and vector Ward
Takahashi identities (WTIs) are satisfied. To this end the quark-gluon interaction in
(\ref{eqn:quarkphotondse}) needs to be the same as the one in the quark DSE.

In general, the quark-photon vertex can be decomposed into twelve covariants
\begin{align}
  \Gamma_\mu(P,k) = \sum_i^{12}\,\,V_\mu^{(i)} \lambda^{(i)}(P,k) \;\;,
  \label{eqn:vertex12Components}
\end{align}
where $V_\mu$ are the covariant tensor structures and $\lambda^{(i)}(P,k)$
are non-trivial scalar dressing functions that contain the
non-perturbative dynamics. The photon momentum is $P$, with $k$ the 
relative quark momentum such that the incoming and outgoing quark 
momenta are $k_\pm = k \pm P/2$. Ball and Chiu suggested to separate
the vertex into the transverse parts $V_\mu^{(5,\ldots,8)}$ with 
$P_\mu V_\mu^{(5,\ldots,8)}=0$ and four non-transverse components 
$V_\mu^{(1,\ldots,4)}$. The latter ones are 
completely fixed in terms of the quark dressing functions $M(p^2)$ 
and $Z_f(p^2)$ by the WTIs and the demand of regularity \cite{Ball:1980ay}. This part of
the quark-photon vertex is also called the Ball-Chiu- or BC-vertex.

The additional eight components of the transverse part are determined 
numerically through a self-consistent solution of the BSE, 
Fig.~\ref{fig:quarkphotondse}. It is an important property of this 
equation that it generates dynamically vector meson bound-state poles.
The idea of vector meson dominance then corresponds to the suggestion
that the transverse part of the quark-photon vertex provides the leading 
contributions in a calculation at hand. While this is indeed correct for
some observables, as for example for $a_\mu^{HVP}$ below, other examples 
are known where this idea is not correct and sizable contributions from 
the BC-part of the vertex occur, see e.g. \cite{Maris:1999ta}. 

\subsection{The hadronic vacuum polarization}
\label{subsec:HVP}

Finally we give some details regarding the calculation
of the hadronic vacuum polarization. Within the truncation scheme
proposed above, the hadronic tensor is given by
\begin{align}
  \Pi_{\mu\nu}(P) =Z_2 \int_q \tr [S(q_-)\Gamma_\mu(P,q)S(q_+)\gamma_\nu ] \;\;,
  \label{eqn:HVPTensor}
\end{align}
where $q_\pm = q \pm P/2$ and
$Z_2$ is the quark wave function renormalisation.
The scalar function $\Pi(P^2)$ is obtained via Eq. (\ref{eqn:HVPScalarDef}).
This quantity is logarithmically divergent and so requires
renormalisation. We apply the condition $\Pi_{\mathrm{R}}(0)=0$ through
the subtraction
\begin{align}
  \Pi_{\mathrm{R}}(P^2):= \Pi(P^2) - \Pi(0) \;\;,
  \label{eqn:RenormalizedHVPscalar}
\end{align}
which effectively amounts to adjusting the constant $Z_3$ in Eq.
(\ref{eqn:photonDSE}) appropriately. In addition we need to take care of 
quadratic divergences that appear through our use of a hard numerical 
cutoff. These can be subtracted at $p^2=0$ or projected out using the 
method of Brown and Pennington \cite{Brown:1988bn}. Both procedures
agree very well.

To check our numerics, we first evaluated the perturbative 
QED one-loop result (see e.g \cite{Nair:2005iw})
and found excellent agreement. In particular we checked that the
calculation of $\Pi_\mathrm{R}$ was independent of the cut-off. As a
further check, we evaluated the electron loop contribution to $a_\mu$ via 
Eq.~(\ref{eqn:anomalyIntegral})
by replacing the propagator and vertices with their tree-level values.
We reproduced the well known result
$a_\mu^{\mathrm{vac. pol.,e-loop}} \approx 5.904\times 10^{-6}$ \cite{Jegerlehner:2009ry} 
on the sub per mille level. For our general calculations with dressed momentum dependent
quark propagator and quark-photon vertex we estimate a numerical error of roughly
two to three percent due to the uncertainties related with the renormalisation 
procedure discussed above. 

\begin{figure}[t]
  \begin{center}
    \includegraphics[width=0.35\textwidth,angle=-90]{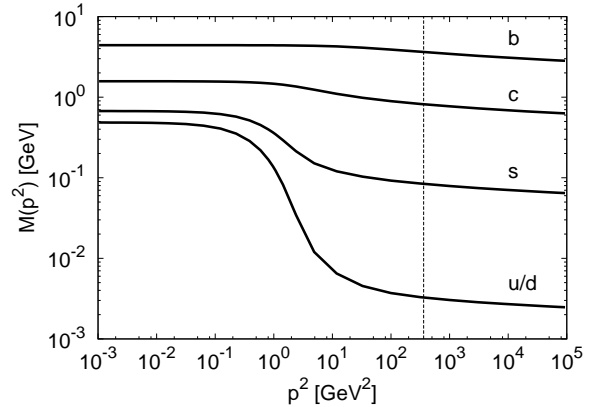}
  \end{center}
  \caption{The quark mass functions of the $u/d$, $s$, $c$ and $b$ quarks obtained from
  the quark DSE. 
  The dashed vertical line represents the renormalization point $\mu^2=(19\,GeV)^2$.}
  \label{fig:QuarkMassFunctions}
\end{figure}

Below we present the results of our calculation for the 
Adler function as well as the anomalous magnetic moment of the muon $a_\mu$.
We use the Maris-Tandy interaction with the two different parameter sets
discussed above. We solve the quark DSE, Fig.~\ref{fig:quarkdse}, for five 
quark flavors $u$, $d$, $s$, $c$ and $b$, and work in the isospin symmetric
limit $m_u=m_d$. The resulting quark mass functions are shown in 
Fig.~\ref{fig:QuarkMassFunctions}.
They dynamically connect the infrared constituent quark 
mass region with the ultraviolet current quark mass region and thus provide 
a unified approach to both pictures. Note that our quark agrees qualitatively 
with lattice calculations~\cite{Bowman:2005vx}.

Once the quarks are obtained we solve for the quark-photon vertex,
Eq.~(\ref{eqn:quarkphotondse}). Here, no additional approximations are made, 
{\it i.e.} we take into account all twelve tensor structures and the full 
momentum dependence of the vertex. This is done for each flavor
separately and hence we can calculate $\Pi_{\mu\nu}$ using Eq. (\ref{eqn:HVPTensor})
which sums over all quark flavors. With $\Pi_{\mu\nu}$ at hand we can
obtain the hadronic contribution to the anomalous magnetic
moment of the muon via Eq.~(\ref{eqn:anomalyIntegral}), and the 
Adler function from Eq.~(\ref{eqn:AdlerFunctionDef}).

\section{Results\label{sec:Results}}
\begin{figure}[t]
  \begin{center}
    \includegraphics[width=0.33\textwidth,angle=-90]{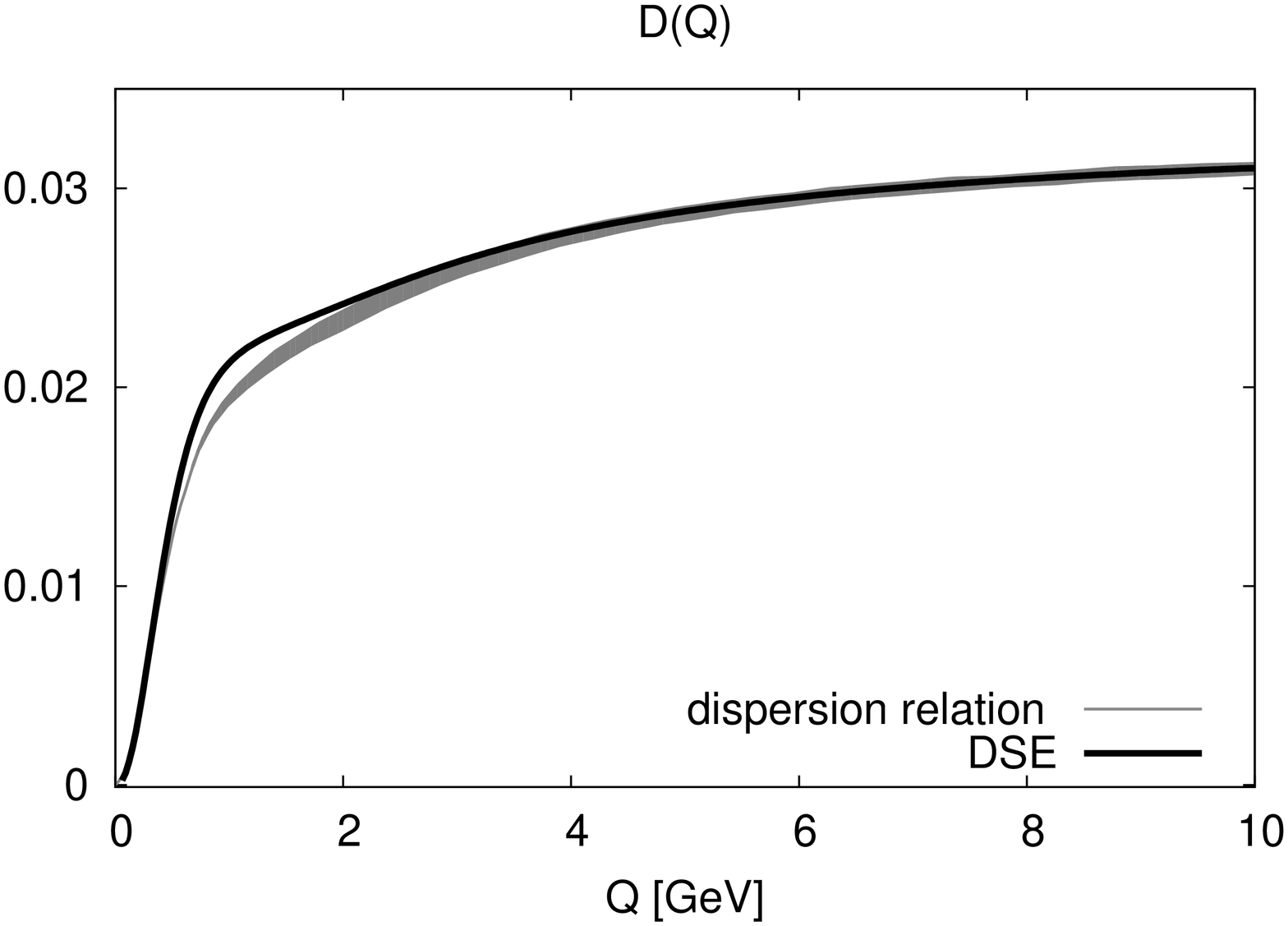}
  \end{center}
  \caption{The Adler function obtained from DSE's for the Maris-Tandy model
   together with the dispersion relation results 
   from~\cite{Eidelman:1998vc,Jegerlehner:2008zza}.\label{fig:Adler}}
%
%
  \begin{center}
    \includegraphics[width=0.33\textwidth,angle=-90]{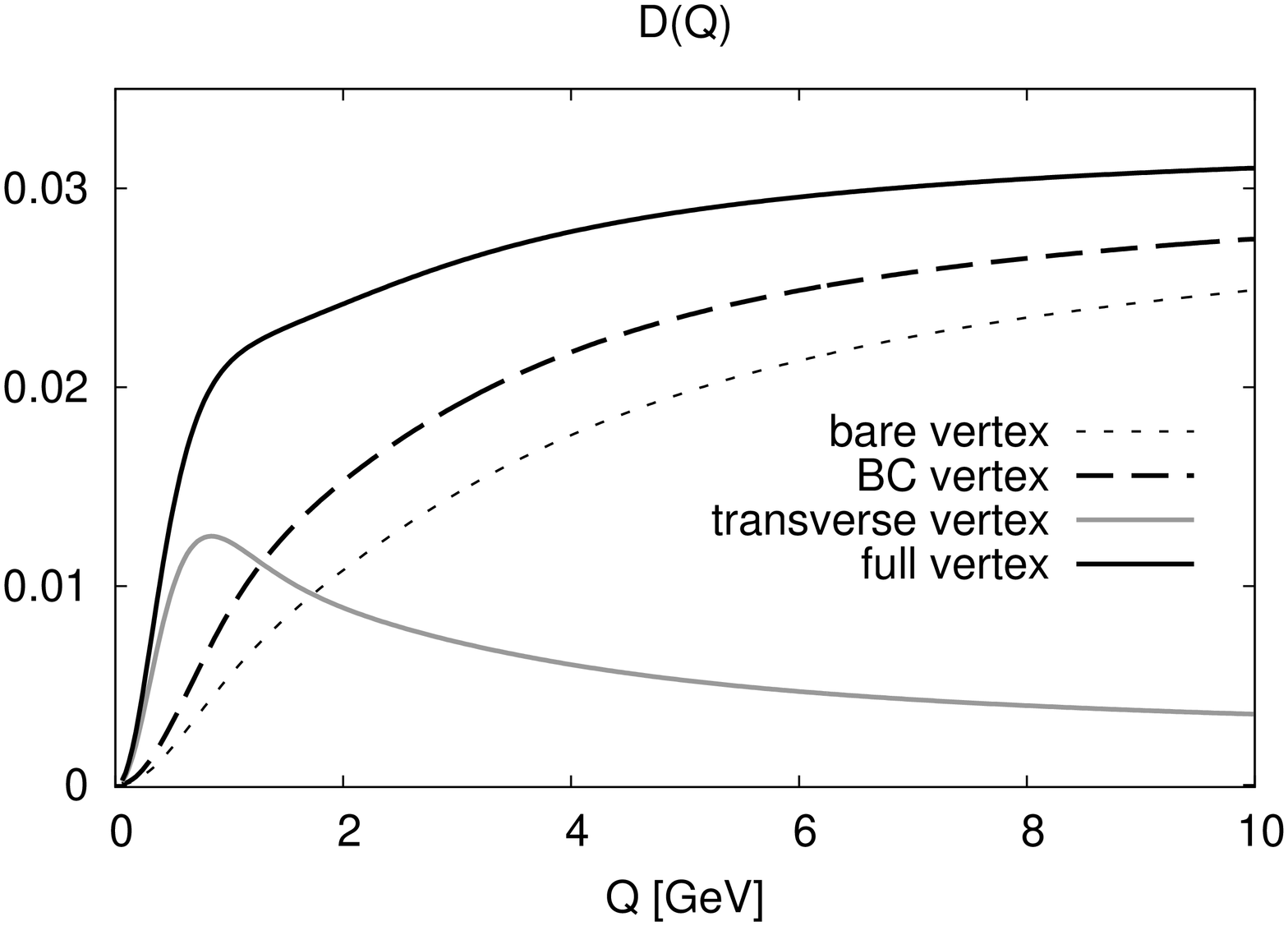}
  \end{center}
  \caption{The Adler function obtained in the MT model defined 
  in Eq. (\ref{eqn:maristandy}) via Eq. (\ref{eqn:HVPTensor}) 
  	with different vertex dressings.}
  \label{fig:AdlerDiffDressings}
\end{figure}

In Fig.~\ref{fig:Adler} we show our result for the Adler function as calculated 
using parameter set II of Table~\ref{tab:param}, together with the result from dispersion 
relations~\cite{Jegerlehner:2009ry,Jegerlehner:2008zza}.
The Dyson--Schwinger solution describes the data very well in the non-perturbative 
region $Q<1\,\,$GeV. We also see that in the asymptotic ultraviolet limit the solution 
follows the result from the dispersion relations. The differences between set I (not shown)
are limited to the slope of the function in the low momentum region, which is most sensitive 
to the mass of the vector meson (see fig. \ref{fig:AdlerDiffDressings}). Note in addition
that most of the contributions to $a_\mu^{HVP}$ come
from the region around the muon mass and that the integration of Eq.~(\ref{eqn:anomalyIntegral})
saturates between $0.5$ and $1$~GeV. From the Adler function we therefore expect
similar results for $a_\mu^{HVP}$ for both parameter sets with small deviations
on the level of ten percent.

Before we discuss our results for $a_\mu^{HVP}$ we take a closer look 
at the impact of the transverse parts of the quark-photon vertex as compared to
its non-transverse Ball-Chiu (BC) structure. In Fig.~\ref{fig:AdlerDiffDressings}
we compare the full results with the one using the BC-part alone or even
neglecting all vertex dressing altogether.
Clearly, the bare and Ball-Chiu vertices do not provide sufficient
contributions to the Adler function, yielding functions that are only
half the height of the full vertex result in the infrared. Only the full vertex 
calculated from its inhomogeneous Bethe-Salpeter equation contains vector 
meson poles dynamically in its transverse structure. Obviously these are 
essential to describe the data correctly. 

This sensitivity to the vector meson sector is especially seen in $a^{HVP}_\mu$. 
For the two mass parameter sets I, II of our model and the full quark-photon vertex we find 
\begin{align}
  a^{{HVP,I}}_\mu &=7440\times 10^{-11}\;\; ,    \\
  a^{{HVP, II}}_\mu & = 6760 \times 10^{-11}\;\;. 
  \label{eqn:amuResults}
\end{align}
As expected, our first mass parameter set yields
a value for $a^{HVP}_\mu$ which is too large by about eight percent, due to
the fact that our vector meson for this parameter set is slightly too light
and can thus be excited from the vaccum too easily. This, however, is
already a reassuringly good result for a calculation
performed with standard parameters without adjustment. Changing our input 
mass parameters to values that are matched to the vector meson sector 
improves our value for $a^{HVP}_\mu$ such that deviations with experiment
fall below three percent. We regard this agreement as a clear signal that our approach 
accurately contains the physics relevant for the hadronic contributions
to $a_\mu$, which entails that indeed the dynamics associated with the vector meson
pole, together with gauge invariance, are the two essential ingredients.

Next we examine the dependence of $a^{{HVP}}_\mu$ on the quark-photon vertex used 
in Eq.~(\ref{eqn:HVPTensor}). The results can be found in Table~\ref{tab:amuResults}.
\begin{table}[t]
	\centering
	\begin{tabular}{c||c|c|c|c}
	$a^{{HVP}}_\mu\times 10^{11}$   &  bare  	&  BC   &  transverse  &  full \\\hline\hline
	set I 							&  $760$	& $1280$ & $6160$        & $7440$\\\hline
	set II 							&  $720$	& $1120$ & $5640$        & $6760$
	\end{tabular}
	\caption{The leading order HVP contribution to $a_\mu$ as obtained by our two
	sets of bare quark masses for different truncations of the quark-photon vertex. 
	\label{tab:amuResults}}
\end{table}
As expected from our results for the Adler function, most of the contribution
to $a_\mu^{HVP}$ comes from the transverse parts of the vertex containing the vector
meson poles. Here also most of the differences between our parameter sets I and II
occur. However, there are also sizable contributions from the gauge or Ball-Chiu part
of the vertex and only the use of the full vertex gives satisfying results for $a_\mu^{HVP}$.
Once more, this emphasizes the interplay of contributions related to resonances
and those demanded by gauge symmetry.

\begin{figure}[t]
  \begin{center}
    \includegraphics[width=0.35\textwidth,angle=-90]{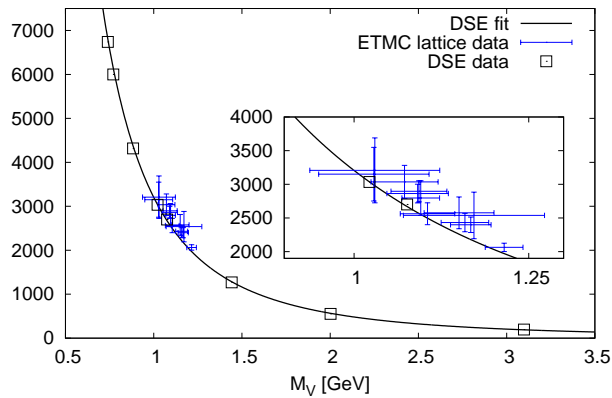}
  \end{center}
  \caption{The mass dependence of $a_\mu^{HVP}\times 10^{11}$, for two flavours,  plotted 
  wrt the mass of the light vector meson. Shown is the data from this
  work (DSE) together with a fit ($a_{\mu}\propto M_V^{(-2.5)}$). In 
  addition we show recent data from the ETMC lattice collaboration~\cite{Feng:2011zk}.}
  \label{fig:a_m}
\end{figure}

Finally we look closer at the dependence of $a_\mu^{HVP}$ on the quark mass. This 
behavior is conveniently parametrized by plotting against a scheme independent, physical 
mass such as for example the pseudoscalar or vector meson mass $m_V$. Both of these can be 
determined in our approach via a solution of their corresponding Bethe-Salpeter equations.
In Fig.~(\ref{fig:a_m}) we show our results for $a_\mu^{HVP}(m_V)$ compared to a 
recent lattice study of the ETMC-collaboration \cite{Feng:2011zk}. Overall we find very good 
agreement between the two approaches, with our values inside their error bars.
The same level of agreement is seen between our calculation and the new lattice determination
for $N_f=2+1$ flavour QCD presented in Ref.~\cite{Boyle:2011hu}.

\section{Discussion} 
Our results for $a_\mu^{HVP}$ clearly show the importance of dressing effects in the 
quark-photon vertex. Here, particularly relevant are its dynamically generated vector 
meson poles in the transverse part of the vertex. However, we wish to emphasize again 
that this importance crucially depends on the kinematic and dynamical details of 
the problem at hand. For example, the transverse parts of the vertex contribute towards 
only half of the pion charge radius \cite{Maris:1999bh}, whilst in the pion pole
approximation of the light by light contributions to g-2 they constitute only a 
thirty percent effect as compared to the BC part~\cite{Fischer:2010iz,Goecke:2010if}.

It is thus very dangerous to transport expectations based on one process blindly to
another; explicit calculations should always be preferred. In this work, we have
performed such a calculation for $a_\mu^{HVP}$ by including both the BC- and transverse
parts of the vertex explicitly. For $a_\mu^{LBL}$ in Refs.~\cite{Fischer:2010iz,Goecke:2010if},
the algebraic complexity forced us to consider initially only the BC part of the vertex, with
transverse parts estimated from other model calculations. Preliminary results for
$a_\mu^{LBL}$ with the full vertex have been presented at~\cite{workshoprw}, and show
that gauge effects still dominate. The details of this will be presented in a future
work.

We also wish to discuss the arguments made in Ref.~\cite{Boughezal:2011vw}. There, 
a constituent quark model with momentum independent masses has been combined with 
a perturbative evaluation of gluonic corrections. Corresponding results for 
$a_\mu^{HVP}$ and $a_\mu^{LBL}$ have been compared. The authors point out that 
neglecting radiative corrections, they need unphysically small constituent quarks
masses to reproduce the experimental value for $a_\mu^{HVP}$. From our results
we can see clearly that this is merely the result of compensating for dynamics that
are absent in the quark-photon interaction of  that model. The authors take note
of that fact and argue that this very light constituent quark mass effectively
includes the $\gamma-\rho$-coupling $g_\rho$ via $M_q\propto M_\rho/g_\rho$. This
simple relation might however be inappropriate for very dissimilar kinematics.
In addition the authors of Ref.~\cite{Boughezal:2011vw}
find very large corrections when they include radiative corrections on the one-loop
level. They observed that these corrections could be absorbed into a
change of the constituent quark mass with stable results for $a_\mu^{HVP}$ 
and $a_\mu^{LBL}$. Based on this result the authors suggest that dressing 
effects in the quark-photon vertex of the full theory should be small. 
We disagree with this conclusion.
First of all, it is dangerous to interpret a truncated perturbative expansion
that features both a large expansion parameter and large expansion coefficients. 
Second, non-perturbative features such as the formation of bound-states (as
generated dynamically by the vertex) are absent in their calculation.
Thus we take their results as a hint that (infinitely many) higher contributions
are important and should be included as consistently as possible, since any finite 
order pQCD cannot give satisfactory answers.

We have done exactly this in our calculation. As a result we found that the leading 
order contribution comes from vector meson (VM) poles accounting for roughly 80 \% 
of $a_\mu^{HVP}$, with the remainder coming from corrections induced by gauge 
invariance. The dynamics of the VM poles are thus important but not the whole story. 
This tells us that an effective model that features only VM exchange should be a good
approximation, but will miss out on other important contributions that cannot be 
integrated by reshuffling of contributions. Similarly, a constituent quark loop approach
would not contain any dynamical degrees of freedom relating to vector meson exchange.
This is in contradiction to what is observed both on the lattice and in our 
Dyson--Schwinger calculation, and thus the constituent quark model cannot be a 
satisfactory description of the process at hand.

Finally, we believe that the good agreement of our results for $a_\mu^{HVP}$ with
experiment \emph{and} with lattice calculations adds credit to our corresponding 
approach to $a_\mu^{LBL}$.

\section{Summary}
\label{sec:Conclusion}
We calculated the hadronic vacuum polarization 
using the method of Dyson-Schwinger equations, taking into account the
five lightest quark flavors. As input we used a phenomenologically successful
model for the quark-gluon interaction together with the rainbow-ladder truncation.
The parameters of these interactions as well as the quark masses were fixed by 
meson observables such as masses and decay constants, without additional fine-tuning. 
We determined the quark-photon vertex from its inhomogeneous Bethe-Salpeter equation 
in the same approximation and subsequently calculated the hadronic vacuum polarization 
tensor. From these we obtained results for the anomalous magnetic moment
of the muon $a^{\mathrm{HVP,LO}}_\mu$ as well as for the Adler function. Both quantities 
agree well with model independent results extracted from experiment. In particular, 
the Adler function is reproduced very well in the strictly non-perturbative region at 
small momenta. We have shown that one requires a description in terms of dynamical quarks 
interacting through non-perturbative gluons in order to achieve this level of accuracy.

Consequently we find results for the muon anomaly in good agreement with other
determinations. Our best result using the quark mass parameter set II is
\begin{align}
  a^{\mathrm{HVP,LO}}_\mu &=6\,760\times 10^{-11}\;\; .
  \label{eqn:amuFinal}
\end{align}
This can be compared to the leading order result quoted in
Eq.~(\ref{eqn:hadroniclo}), $6\,903.0(52.6)\times 10^{-11}$. The difference
is at the level of two percent. A comparison with the result
$a^{\mathrm{HVP,LO}}_\mu =7\,440\times 10^{-11}$ obtained with our parameter 
set I may serve as an estimate for the systematic uncertainty of our model of 
roughly ten percent. We believe our approach to the hadronic light-by-light scattering
contribution~\cite{Fischer:2010iz,Goecke:2010if}, which employs the same
truncation scheme, will ultimately lead to results with similar precision. 
However, note that in Ref.~\cite{Fischer:2010iz,Goecke:2010if} the full quark-photon
vertex was not yet included in the quark-loop due to its algebraic
complexity. Improvements along this direction are underway.

\section{Acknowledgments}  
We thank A.~E.~Dorokhov and A.~E.~Radzhabov for helpful discussions. 
This work was supported by the DFG under grant No.~Fi 970/8-1, by the 
Helmholtz-University Young Investigator Grant No.~VH-NG-332 and by the 
Helmholtz International Center for FAIR within the LOEWE program of the 
State of Hesse. 
RW would also like to acknowledge support by the Austrian
Science Fund FWF under Project No. P20592-N16, and by 
Ministerio de Educaci\'on (Spain): Programa Nacional de Movilidad de 
Recursos Humanos del, Plan Nacional de I-D+i 2008-2011.

\end{document}